\documentclass[referee,a4paper,12pt,traditabstract]{jswsc} 

\usepackage{graphicx}
\usepackage{txfonts}
\usepackage{subfigure}
\usepackage{epstopdf}
\usepackage[mathlines]{lineno}
\usepackage[authoryear,round]{natbib}
\usepackage[backref]{hyperref}
\usepackage{url}

\usepackage{array,booktabs,threeparttable}
\newcolumntype{C}{>{$\displaystyle}c<{$}}

\hypersetup{colorlinks=true,citecolor=blue,urlcolor=magenta,linkcolor=magenta}

\usepackage{journal_macros}

\graphicspath{{./}{figures/}}

\usepackage[usenames,dvipsnames]{xcolor}
\definecolor{cadmiumgreen}{rgb}{0.0, 0.42, 0.24}
\newcommand\editone[1]{\textcolor{black}{#1}}
\newcommand\edittwo[1]{\textcolor{black}{#1}}

\begin{document}


   \title{\edittwo{Improved modelling of SEP event onset within the WSA--Enlil--SEPMOD framework}}
   
   \titlerunning{Improving SEP event onsets with SEPMOD}

   \authorrunning{Palmerio et al.}

   \author{
   Erika~Palmerio\inst{1}
   \and
   Janet~G.~Luhmann\inst{2}
   \and
   M.~Leila~Mays\inst{3}
   \and
   Ronald~M.~Caplan\inst{1}
   \and
   David~Lario\inst{3}
   \and
   Ian~G.~Richardson\inst{3,4}
   \and
   Kathryn~Whitman\inst{5,6}
   \and
   Christina~O.~Lee\inst{2}
   \and
   Beatriz~S{\'a}nchez-Cano\inst{7}
   \and
   Nicolas~Wijsen\inst{8}
   \and
   Yan~Li\inst{2}   
   \and
   Carlota~Cardoso\inst{9,10}
   \and
   Marco~Pinto\inst{11,9}
   \and
   Daniel~Heyner\inst{12}
   \and
   Daniel~Schmid\inst{13}
   \and
   Hans-Ulrich~Auster\inst{12}
   \and
   David~Fischer\inst{13}
   }

   \institute{
   Predictive Science Inc., San Diego, CA 92121, USA\\
   \email{\href{mailto:epalmerio@predsci.com}{epalmerio@predsci.com}}
   \and
   Space Sciences Laboratory, University of California--Berkeley, Berkeley, CA 94720, USA
   \and
   Heliophysics Science Division, NASA Goddard Space Flight Center, Greenbelt, MD 20771, USA
   \and
   Department of Astronomy, University of Maryland, College Park, MD 20742, USA
   \and
   University of Houston, Houston, TX 77204, USA
   \and
   KBR, Inc., Houston, TX 77002, USA
   \and
   School of Physics and Astronomy, University of Leicester, Leicester, LE1 7RH, UK
   \and
   Department of Mathematics/Centre for Mathematical Plasma Astrophysics, KU Leuven, B-3001 Leuven, Belgium
   \and
   Laborat{\'o}rio de Instrumenta\c{c}\~{a}o e F{\'i}sica Experimental de Part{\'i}culas, 1649-003 Lisbon, Portugal
   \and
   Instituto Superior T{\'e}cnico, University of Lisbon, 1049-001 Lisbon, Portugal
   \and 
   European Space Research and Technology Centre, European Space Agency, 2201 AZ Noordwijk, The Netherlands
   \and
   Institut f{\"u}r Geophysik und extraterrestrische Physik, TU Braunschweig, D-38106 Braunschweig, Germany
   \and
   Space Research Institute, Austrian Academy of Sciences, A-8042 Graz, Austria
    }


 
  \abstract
   {
\edittwo{Multi-spacecraft observations of solar energetic particle (SEP) events not only enable a deeper understanding and development of particle acceleration and transport theories, but also provide important constraints for model validation efforts.} However, because of computational limitations, a given physics-based SEP model is usually best-suited to capture a particular phase of an SEP event, rather than its whole development from onset through decay. For example, magnetohydrodynamic (MHD) models of the heliosphere often incorporate solar transients only at the outer boundary of their so-called coronal domain---usually set at a heliocentric distance of 20--30\,$R_{\odot}$. This means that particle acceleration at CME-driven shocks is also computed from this boundary onwards, leading to simulated SEP event onsets that can be many hours later than observed, since shock waves can form much lower in the solar corona. In this work, we aim to improve the modelled onset of SEP events by inserting a ``fixed source'' of particle injection at the outer boundary of the coronal domain of the coupled WSA--Enlil 3D MHD model of the heliosphere. The SEP model that we employ for this effort is SEPMOD, a physics-based test-particle code based on a field line tracer and adiabatic invariant conservation. We apply our initial tests and results of SEPMOD's fixed-source option to the 2021 October 9 SEP event, which was detected at five well-separated locations in the inner heliosphere---Parker Solar Probe, STEREO-A, Solar Orbiter, BepiColombo, and near-Earth spacecraft.
   }
   

   \maketitle


\section{Introduction} \label{sec:intro}

Solar energetic particle (SEP) events can display a wide variation in their properties such as intensity, fluence, composition, temporal profile, energy spectrum, and spatial extent. Traditionally, their acceleration sites have been distinguished between solar flares, giving rise to impulsive SEP events \citep[e.g.,][]{lin2011}, and coronal mass ejection (CME)-driven shocks, thus generating gradual SEP events \citep[e.g.,][]{desai2016}. Whilst impulsive SEP events are characterised by a prompt onset and a rapid decay, resulting in a short duration (order of a few hours), gradual ones tend to feature variable profiles (depending on the observer's location with respect to the shock nose), can last up to a few days, and may include a local energetic storm particle (ESP) event in the vicinity of the CME-driven shock passage. Nowadays, this dichotomy has become more blurred, as both flares and CMEs have been shown to be able to contribute to SEP acceleration within a single event \citep[e.g.,][]{cane2010, trottet2015}. Significant progress in understanding SEP acceleration and transport has been achieved owing to multi-point measurements, which have allowed the study of, e.g., the broad longitudinal distribution of so-called widespread events \citep[e.g.,][]{lario2014b, xie2017, kollhoff2021, dresing2023}, how the local heliospheric conditions may facilitate or hinder the transport of particles up to a given location \citep[e.g.,][]{richardson1996, lario2014a, bain2016, palmerio2021}, as well as the longitudinal and/or radial variation of SEP profiles and properties \citep[e.g.,][]{rouillard2012, cohen2014, lario2016, wellbrock2022}. In addition to providing a wealth of data for the development and refinement of theories of particle acceleration and transport, multi-point observations of SEPs are also crucial for validating modelling efforts, since they conveniently provide results at multiple locations that can test the predictions of a single simulation run.

The current status of SEP modelling has been recently reviewed by \citet{whitman2023}, who highlighted that the models that are presently existing or in development employ a wide variety of approaches and have different goals. For example, empirical models are based on correlations found between observational parameters and the properties of the subsequent SEP event that may be related to one or more underlying physical processes \citep[e.g.,][]{posner2007, anastasiadis2017, bruno2021}. They are able to provide rapid forecasts, often in the form of an ``all clear'' or ``not all clear'' prediction, representing e.g.\ the confidence level that an SEP event of a certain magnitude will be observed at a given location. A somewhat similar approach is undertaken by more recently-developed machine learning models \citep[e.g.,][]{aminalragiagiamini2021, lavasa2021, kasapis2022}, which are usually trained with large data sets and thus aim to provide prompt forecasts with improved accuracy, and as a consequence by ``hybrid'' models that combine empirical and machine learning characteristics \citep[e.g.,][]{nunez2017, laurenza2018}. Another class of SEP models is represented by physics-based ones, which make use of our current understanding of particle acceleration and transport to simulate the full distribution of SEPs from a certain event in 2D or 3D space \citep[e.g.,][]{schwadron2010, zhang2017, tenishev2021}. Some of these models can be coupled to magnetohydrodynamic (MHD) models of the corona and/or heliosphere, yielding thus a self-consistent description of particle acceleration at e.g.\ a CME-driven shock as it propagates away from the Sun \citep[e.g.,][]{sokolov2004, linker2019, li2021}. Despite their generally more realistic representation of the physics at play, most physics-based models are computationally intensive and may require estimates of physical parameters that are poorly determined from observations or theory, and thus may not be best suited for real-time forecasts with current high-performance computing capabilities and/or costs.

Another issue found when coupling physics-based \editone{SEP transport and MHD heliospheric models} is that, because of computational constraints, MHD models are usually separated into two spatial domains: A coronal one (generally extending up to 20--30\,$R_{\odot}$) and a heliospheric one (typically covering the region from 20--30\,$R_{\odot}$ to 1~au and beyond). In most cases, the coronal and heliospheric domains are characterised not only by different spatial scales, but also slightly different physical assumptions. A subset of these models, in particular, circumvent the complex modelling of coronal dynamics by inserting solar transients (such as CMEs) only at the heliospheric inner boundary. This simplified (and faster) approach has been shown to be adequate when simulating, e.g., CME arrival time \citep[e.g.,][]{lee2015, wold2018}, flux rope properties \citep[e.g.,][]{scolini2019, maharana2022}, and/or ESP events \citep[e.g.,][]{ding2022, wijsen2022} in interplanetary space. However, \editone{neglecting solar transients below the heliospheric inner boundary} inevitably results in missing the critical early acceleration phase of SEPs \citep[e.g.,][]{palmerio2022}, since both solar flares and CME-driven shocks are able to produce and release energetic particles at much lower coronal heights \citep[e.g.,][]{kozarev2015, masson2019, young2021}.

\begin{figure}[t!]
\centering
\includegraphics[width=.5\linewidth]{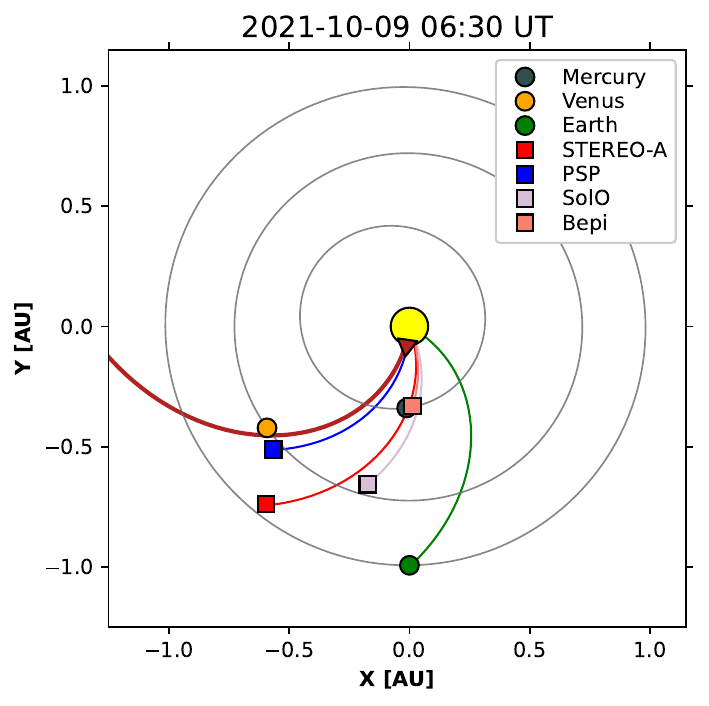}
\caption{Position of planets and spacecraft within 1~AU from the Sun on 2021 October 9 at 06:30~UT. The source region of the CME eruption associated with an M1.6-class flare is indicated with a triangle symbol on the surface of the Sun. Earth and the spacecraft have been connected to the Sun through the nominal Parker spiral for a solar wind speed of 400~km$\cdot$s$^{-1}$. The same spiral connects the source region of the eruption to the heliosphere. The orbits of Mercury, Venus, and Earth are also shown.}
\label{fig:orbits}
\end{figure}

In this work, we aim to mitigate this issue by ``artificially'' inserting sources that inject SEPs at the inner heliospheric boundary of one such 3D MHD model. The modelling suite that we employ consists of the coronal Wang--Sheeley--Arge \citep[WSA;][]{arge2000, arge2004}, the heliospheric Enlil \citep{odstrcil2003, odstrcil2004}, and the Solar Energetic Particle MODel \citep[SEPMOD;][]{luhmann2007, luhmann2010} codes. We apply the model architecture to the 2021 October 9 SEP event, which was observed at five locations in the inner heliosphere that were relatively well connected to the eruption's source region (see Figure~\ref{fig:orbits}). Several studies have addressed some aspects of this event. For example, \citet{lario2022} analysed the large-scale structure of the interplanetary context and studied its influence on the transport of SEPs to each observer. \citet{yang2023} investigated the evolution of the CME associated with the SEP event and its interaction with the structured solar wind. \citet{jebaraj2023} examined the acceleration of electrons by both the solar flare temporally associated with the parent solar eruption and by the shock driven by the corresponding CME. Finally, \citet{wijsen2023} simulated the propagation of the CME-driven shock and the ESP event at two of the five observers using the European Heliospheric FORecasting Information Asset \citep[EUHFORIA;][]{pomoell2018} and PArticle Radiation Asset Directed at Interplanetary Space Exploration \citep[PARADISE;][]{wijsen2019} models.

After providing an overview of the 2021 October 9 eruption (Section~\ref{sec:erupt}), we first show modelling results with WSA--Enlil--SEPMOD in their ``default'' settings and compare them with interplanetary magnetic field, solar wind plasma, and particle data as supplied by the spacecraft at the five locations (Section~\ref{subsec:default}). We then detail the implementation of the so-called ``fixed-source'' option in SEPMOD (Section~\ref{subsec:fixedsource}) by simulating two different particle acceleration sites: one connected to the active region from which the eruption originated (``solar flare source'') and the other related to the CME-driven shock in the corona (``coronal shock source''). We compare results from both fixed-source options with SEP measurements at the five observers (Section~\ref{sec:result}), expecting to find improvements in the simulated event onset time(s) with respect to the default model architecture. Finally, we discuss our results in the context of space weather predictions of SEP events (Section~\ref{sec:discuss}) and provide concluding thoughts as well as recommendations for future improvements in physics-based particle modelling (Section~\ref{sec:conclu}).


\section{Overview of the 2021 October 9 eruption} \label{sec:erupt}

\begin{figure}[th!]
\centering
\includegraphics[width=.99\linewidth]{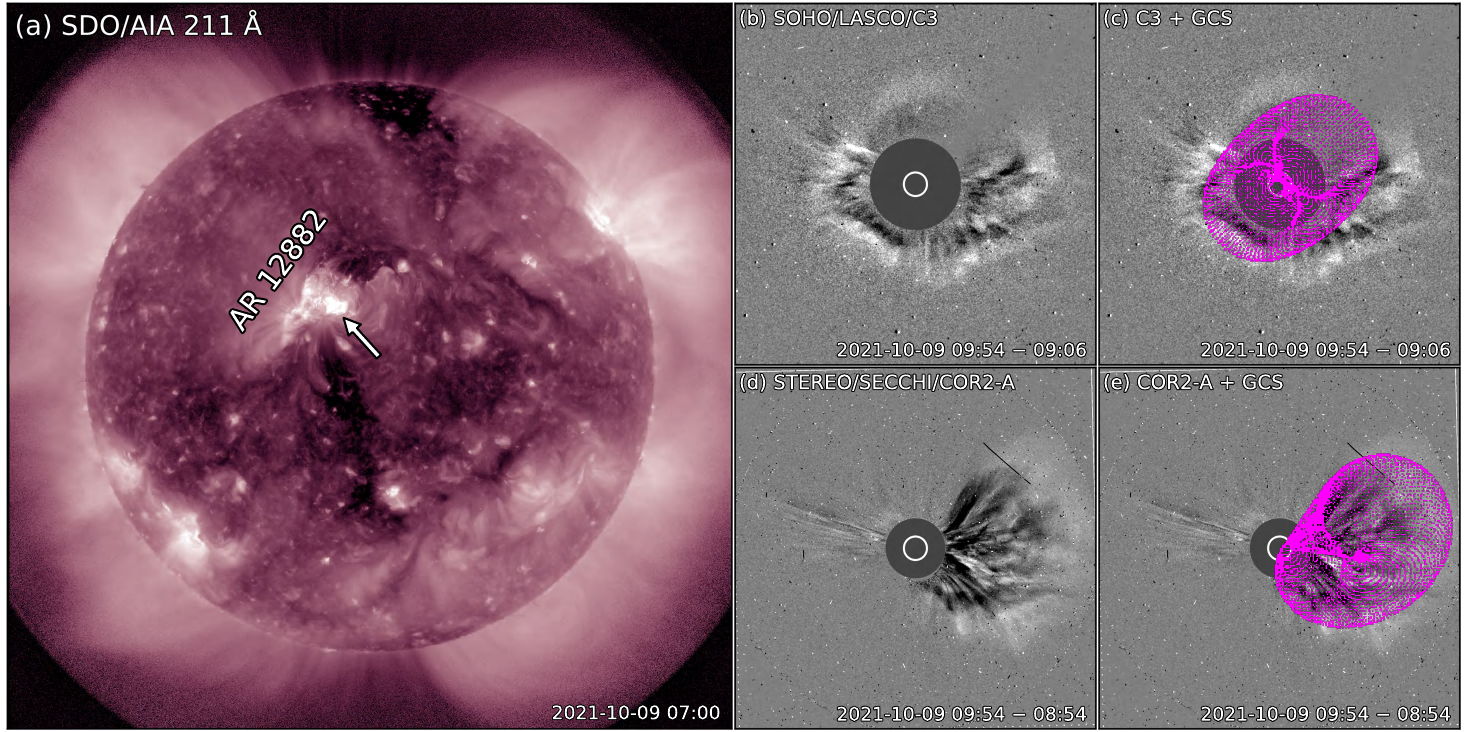}
\caption{Overview of the 2021 October 9 CME eruption from remote-sensing imagery. (a) Image in the 211~{\AA} channel taken by SDO/AIA showing AR 12882 erupting. (b) Coronagraph difference image from SOHO/LASCO/C3, shown in (c) with the GCS wireframe overlaid. (d) Coronagraph difference image from STEREO/SECCHI/COR2-A, shown in (e) with the GCS wireframe overlaid.}
\label{fig:remotesens}
\end{figure}

The event that we focus on in this work commenced on 2021 October 9 around 06:30~UT. It originated from NOAA active region (AR) 12882, located at N18E08 (in Stonyhurst coordinates) at the time of the eruption and highlighted in Figure~\ref{fig:remotesens}(a), which displays data from the Atmospheric Imaging Assembly \citep[AIA;][]{lemen2012} telescope onboard the Earth-orbiting Solar Dynamics Observatory \citep[SDO;][]{pesnell2012}. The eruption was associated with a CME, seen to move slightly to the northwest of its source region in SDO/AIA imagery, an extreme ultra-violet (EUV) wave \citep[described in detail by][]{lario2022}, and an M1.6-class flare with start time 06:19~UT, peak time 06:38~UT, and end time 06:53~UT as observed by the Geostationary Operational Environmental Satellite (GOES). The northwesterly motion of the CME and associated EUV wave may be due to the presence of coronal holes to the south of AR 12882, as shown in Figure~\ref{fig:remotesens}(a).

The CME propagation through the solar corona was observed in white light from two perspectives, namely the Solar and Heliospheric Observatory \citep[SOHO;][]{domingo1995}, placed at the Sun--Earth L1 point, and the Solar Terrestrial Relations Observatory Ahead \citep[STEREO-A;][]{kaiser2008} spacecraft, located ${\sim}$40$^{\circ}$ east of Earth and about 1~au from the Sun at the time of the event. Coronagraph imagery from these two observers is shown in Figure~\ref{fig:remotesens}(b)--(e), which displays data from the C3 telescope of the Large Angle and Spectrometric Coronagraph \citep[LASCO;][]{brueckner1995} onboard SOHO and the COR2 telescope of the Sun Earth Connection Coronal and Heliospheric Investigation \citep[SECCHI;][]{howard2008} onboard STEREO-A. It is clear from these data that the CME appeared as a full halo from Earth's perspective, whilst its angular width from the STEREO-A viewpoint was ${\sim}$130$^{\circ}$. These observations suggest that the 2021 October 9 CME was Earth-directed, with the possibility of a flank encounter at most at STEREO-A.

To obtain a more quantitative estimate of the CME size, trajectory, and speed (especially useful for the heliospheric modelling described in Section~\ref{sec:model}), we perform reconstructions of the eruption in the corona using the Graduated Cylindrical Shell \citep[GCS;][]{thernisien2011} model applied to nearly-simultaneous SOHO/LASCO and STEREO/SECCHI images. An example of the GCS wireframe---basically a hollow croissant-like shell described by six free parameters---is projected onto coronagraph data in  Figure~\ref{fig:remotesens}(c) (SOHO) and Figure~\ref{fig:remotesens}(e) (STEREO-A). According to our reconstruction(s), the CME nose propagated in the direction ($\theta$, $\phi$) = (9$^{\circ}$, 5$^{\circ}$) in Stonyhurst coordinates, and its axis was tilted by $\gamma$ = 40$^{\circ}$ counterclockwise to the western direction at the solar equator. We obtain the CME angular extent both along its axis and in the direction perpendicular to it by calculating the semi-angular widths of the projection of the GCS wireframe as seen ``from its top'' \citep[approximately, the ellipse that is visible in Figure~\ref{fig:remotesens}(c); see also][]{verbeke2023}, resulting in ($R_\mathrm{maj}$, $R_\mathrm{min}$) = (49$^{\circ}$, 28$^{\circ}$). The CME speed is estimated by performing two GCS reconstructions at times separated by one hour and by considering the radial distance travelled by the CME nose in between, resulting in $V$ = 770~km$\cdot$s$^{-1}$ (calculated when the CME had reached heights in the range ${\sim}$12--17\,$R_{\odot}$). Thus, through the solar corona, the 2021 October 9 CME travelled close to the Sun--Earth line, spanned ${\sim}$100$^{\circ}$ along its axis, and was characterised by a moderate speed.


\section{Modelling the 2021 October 9 CME and SEP event} \label{sec:model}

Here, we focus on modelling the 2021 October 9 CME and SEP event using the combined WSA--Enlil--SEPMOD architecture. WSA is a model that operates within the simulation's coronal domain, i.e.\ in the range 1--21.5\,$R_{\odot}$ \editone{from the centre of the Sun}. It employs photospheric magnetic field maps of the full Sun to generate solar wind and interplanetary magnetic field conditions up to 21.5\,$R_{\odot}$ (or 0.1~au), corresponding to the WSA outer boundary and also the Enlil inner boundary. Enlil then uses the WSA output as input to model the conditions within the simulation's heliospheric domain (from 0.1~au onwards) by solving the MHD equations via a flux-corrected-transport algorithm. In this work, we set the outer boundary of the simulation to 2~au. CMEs are directly inserted at the Enlil inner boundary of 0.1~au as hydrodynamic pulses, i.e.\ without an internal magnetic field. Enlil results are then read by the SEPMOD code to generate SEP fluxes (typically, for protons) at the observer(s) of interest. Specifically, it forward-models the transport of SEPs along field lines that connect a CME-driven shock to a specific observer, using shock-source injections followed by a constant-energy, guiding-center transport approximation---it follows that the transport of particles is assumed to be scatter-free, i.e.\ diffusive propagation is neglected. In this section, we first describe the modelling set up using the default WSA--Enlil--SEPMOD architecture (Section~\ref{subsec:default}), and then we illustrate the implementation of the fixed-source option in SEPMOD to emulate the early-acceleration phase of SEP events (Section~\ref{subsec:fixedsource}).

\subsection{Using the default WSA--Enlil--SEPMOD architecture} \label{subsec:default}

\begin{figure}[th!]
\centering
\includegraphics[width=.99\linewidth]{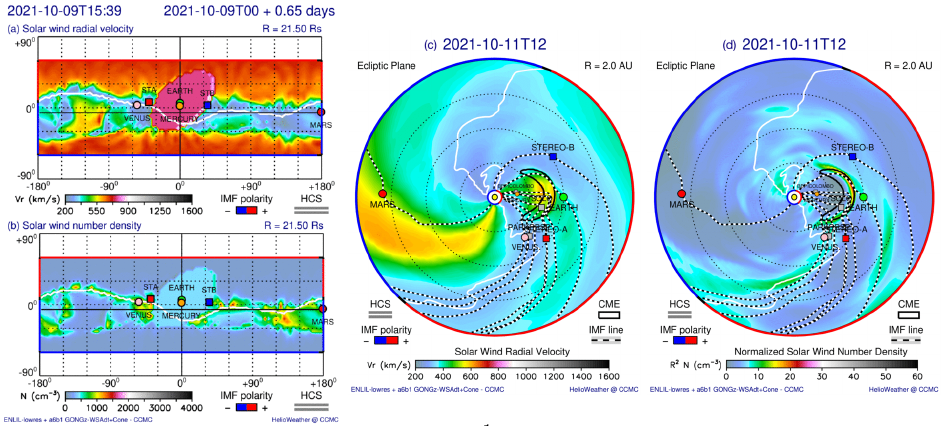}
\caption{Overview of the WSA--Enlil simulation of the 2021 October 9 CME. (a) Radial slice of the solar wind radial speed at the Enlil inner boundary of 21.5\,$R_{\odot}$, showing the injection of the CME at its point of maximum width. (b) Same as (a), but for the solar wind number density. (c) View on the ecliptic plane of the solar wind radial speed within the simulation's heliospheric domain (0.1--2.0~au), showing the CME approaching Earth. (d) Same as (c), but for the normalised solar wind number density. The CME ejecta is traced from its injection based on the density enhancement with respect to the background solar wind (initially set to 4 times the ambient density), and is outlined with a black contour in panels (c) and (d).}
\label{fig:enlil_setup}
\end{figure}

The first step in our modelling efforts consists of setting up the WSA--Enlil--SEPMOD simulation run that only uses CME-driven shock information in the heliospheric domain to produce particles---we will refer to this default set up as the ``interplanetary shock source'' for SEPMOD. This run is performed at NASA's Community Coordinated Modeling Center (CCMC). As input conditions for WSA, we use time-dependent (hourly) National Solar Observatory (NSO) Global Oscillation Network Group \citep[GONG;][]{harvey1996} zero-point-corrected magnetogram synoptic maps. The CME input parameters are derived directly from the GCS reconstructions presented in Section~\ref{sec:erupt}, and the time of injection into Enlil is derived from the CME speed calculated as far into the COR2-A field of view as possible and propagated to 21.5\,$R_{\odot}$ under the assumption of constant velocity. This results in the CME being injected into the heliospheric domain at $t_{0}$ = 2021-10-09T11:03, with nose trajectory ($\theta$, $\phi$) = (9$^{\circ}$, 5$^{\circ}$) in Stonyhurst coordinates, tilt $\gamma$ = 40$^{\circ}$ (defined positive for counterclockwise rotations from the west direction), half-angular widths ($R_\mathrm{maj}$, $R_\mathrm{min}$) = (49$^{\circ}$, 28$^{\circ}$), and initial speed $V$ = 770~km$\cdot$s$^{-1}$. Note that these input parameters result in a CME morphology that consists of a tilted ellipsoid rather than the classic (spherical) ice-cream cone \citep[see also][]{mays2015}. An overview of the WSA--Enlil simulation, with the CME being inserted at the 21.5\,$R_{\odot}$ boundary and its propagation through the inner heliosphere, is shown in Figure~\ref{fig:enlil_setup}.

Once the WSA--Enlil run is complete, SEPMOD uses shock and connectivity information from Enlil to model SEP fluxes at the observers of interest. In Enlil, shocks are defined as increases in the solar wind speed by ${\geq}$20~km$\cdot$s$^{-1}$ with respect to the corresponding ambient simulation (i.e., the run without CMEs), and the shock speed is computed as a time-derivative of the shock position. In SEPMOD, the particle source at the interplanetary shock is described by a time sequence of point sources on each observer-connected field line where a shock is found. The Enlil-derived shock parameters (generally speed, density and speed jumps, as well as shock location) are used, together with the observer-connected field lines, to model the injection of protons at the time of  each connection. Their subsequent trajectories along the field line are traced under the guiding-centre approximation, integrating as the observer passes from field line to field line to obtain the ``observed'' event flux time profile. The default energy distribution of each proton injection is an inverse power-law whose spectral index ($\alpha$) depends on the Enlil shock compression ratio at the point that is magnetically connected to the observer, according to the standard formula for diffusive shock acceleration \citep[e.g.,][]{jones1991}:
\begin{equation}
\alpha = -\frac{1}{2}\frac{J_{n} + 2}{J_{n} -1} \, ,
\label{eq:spectralindex}
\end{equation}
where $J_{n}$ is the density jump at the shock, whilst its intensity at a certain energy is determined by the velocity jump using empirical results from \citet{lario1998}. The default injected pitch angle distribution is assumed to be isotropic.

\subsection{Introducing the fixed-source option in SEPMOD} \label{subsec:fixedsource}

The fixed-source option in SEPMOD aims to mitigate the lack of simulated solar transients---and, thus, of CME-driven shocks and SEP injection regions---within the coronal domain of models such as WSA--Enlil. The scheme is realised by setting a predefined source of particles at the coronal--heliospheric model boundary and by regulating the corresponding injection by means of properties such as width of the source, timing of the injection, energy spectrum, and decay profile. A preliminary version of this concept was described by \citet{luhmann2012}, who employed \editone{ad-hoc} values to model the early onset of a series of SEP events in 2010 August. In this work, we expand upon the initial implementation of \citet{luhmann2012} to use insights from more recent work to describe the fixed source, as well as to combine the resulting SEP fluxes with those that emerge later on from the interplanetary shock source once the CME has been inserted at the heliospheric inner boundary of the simulation. Furthermore, here we consider two separate fixed-source descriptions, to emulate the two possible particle acceleration sites \citep[e.g.,][]{reames1999}:

\begin{itemize}
\item \textbf{Solar flare source:} This source is directly related to the active region where an eruption originates from and approximates the flare acceleration site---very close to the solar surface---that gives rise to so-called impulsive SEP events.
\item \textbf{Coronal shock source:} This source is directly related to the early propagation of the corresponding CME and approximates the low-coronal shock acceleration site that gives rise to so-called gradual SEP events---as the CME travels away from the Sun, this source naturally morphs into the interplanetary shock source.
\end{itemize}

The criteria that we define for each of the two fixed-source descriptions are summarised in Table~\ref{tab:fixedsource}, and further details are provided throughout the rest of this section. The solar flare source option uses separate Potential Field Source Surface \citep[PFSS;][]{wang1992} mapping of the flare site to the source surface to infer the location and area of the fixed source at 21.5\,$R_{\odot}$, where it assumes protons are injected with an initial peak flux weighted by an empirical factor depending on the 1--8~{\AA} X-ray intensity and that \editone{falls exponentially with a decay constant that is scaled with the corresponding flare class}. In contrast, the coronal shock source is assumed to be located on the 21.5\,$R_{\odot}$ Enlil inner boundary at the location of the cone CME insertion, where it assumes injection of protons over the area of the cone CME with an initial rollover spectrum (with rollover energy set to 10~MeV) that evolves to a power law with spectral index $\alpha = -$2. The assumed exponential decay time constant for this source is currently the transit time out to 21.5\,$R_{\odot}$ based on the CME speed in the corona.

\begin{table}[t!]
\caption{Criteria used to define the solar flare and coronal shock SEP sources. \label{tab:fixedsource} \vspace*{.1in}}
\centering
\renewcommand{\arraystretch}{1.3}
\begin{tabular}{l@{\hskip .3in}c@{\hskip .3in}c}
\toprule
 & \textsc{\textbf{Solar Flare Source}} & \textsc{\textbf{Coronal Shock Source}} \\
\midrule
Spatial location & AR open field at 2.5\,$R_{\odot}$ & Maximum CME width at ${\sim}$10--15\,$R_{\odot}$\\
Injection time & X-ray flare peak & Type II onset\\
Energy spectrum & Steep/soft power law & Rollover spectrum\\
Maximum intensity & Scaled with flare class & Scaled with CME speed\\
Temporal profile & Exponential decay from flare time & Exponential decay from Type Il time\\
Decay constant & Scaled by flare magnitude & Based on CME transit time to 21.5\,$R_{\odot}$\\
\bottomrule
\end{tabular}
\vspace*{.1in}
\begin{tablenotes}
\item \emph{Notes.} See the text in Section~\ref{subsec:fixedsource} for further details.
\end{tablenotes}
\end{table}

\begin{figure}[th!]
\centering
\includegraphics[width=.99\linewidth]{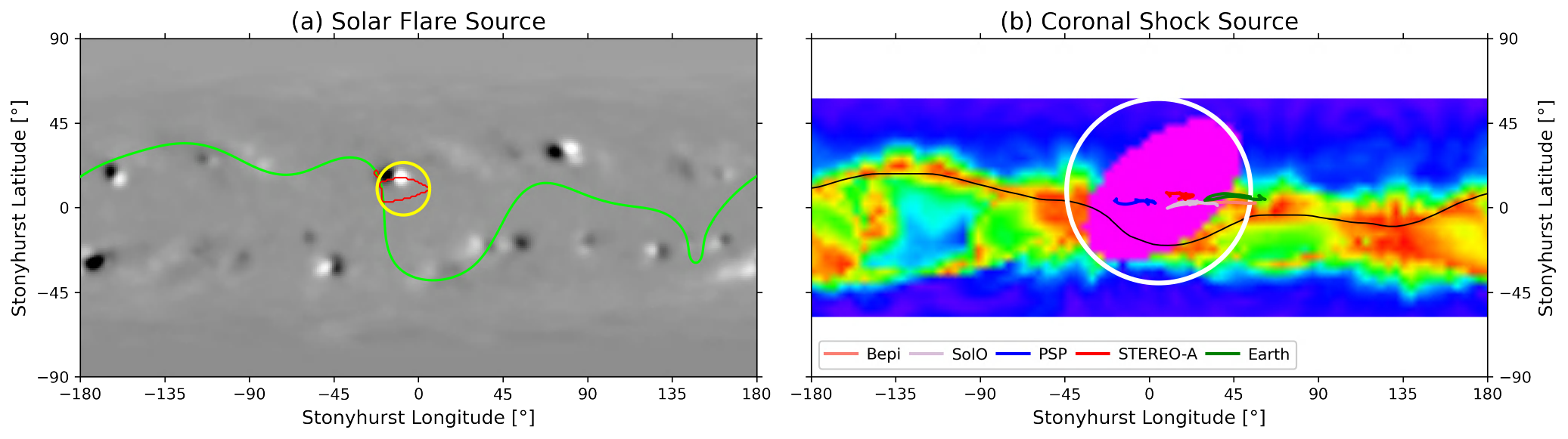}
\caption{Overview of the selected fixed sources for the 2021 October 9 event. (a) Solar flare source. The panel shows the smoothed GONG zero-point-corrected magnetogram for 2021-10-09 at 06:04~UT (saturated at ${\pm}$50~G), the PFSS-derived location of the heliospheric current sheet at 2.5\,$R_{\odot}$ (lime contour), the mapping of AR~12882 onto the 2.5\,$R_{\odot}$ PFSS source surface (red contour), and the selected fixed-source area to employ for SEP injection (yellow contour). (b) Coronal shock source. The panel shows the solar wind radial speed at the Enlil inner boundary of 21.5\,$R_{\odot}$ at the point of maximum width of the CME insertion (2021-10-09 at 15:39~UT, same as Figure~\ref{fig:enlil_setup}(a)), the location of the modelled heliospheric current sheet (black contour), the selected fixed-source area to employ for SEP injection (white contour), and the projected Enlil-modelled field lines connecting each of the five observers to the 21.5\,$R_{\odot}$ sphere.}
\label{fig:fixed_sources}
\end{figure}

An overview of the spatial locations and extents of the two fixed sources selected for the 2021 October 9 event is presented in Figure~\ref{fig:fixed_sources}. The extent of the solar flare source (Figure~\ref{fig:fixed_sources}(a)) is obtained from the GONG magnetogram that is closest to the eruption time from the series of input maps used for WSA (see Section~\ref{subsec:default}), i.e.\ the one for October 9 at 06:04~UT. After processing the magnetogram via interpolation, flux balancing, and smoothing, we perform on it a PFSS extrapolation using the POT3D \citep{caplan2021} tool. We then select a mask of AR~12882 (by thresholding it to ${\pm}$10~G) and trace field lines from it onto the 2.5\,$R_{\odot}$ PFSS source surface (represented with a red contour in Figure~\ref{fig:fixed_sources}(a)). Finally, we select a circular area encompassing the mapped AR to represent the solar flare source---shown in yellow in Figure~\ref{fig:fixed_sources}(a)---resulting in a patch centred at ($\theta$, $\phi$) = ($10^{\circ}$, $-8^{\circ}$)  in Stonyhurst coordinates and with a $14^{\circ}$ radius. The extent of the coronal shock source (Figure~\ref{fig:fixed_sources}(b)) is obtained from the GCS-derived (see Figure~\ref{fig:remotesens}) CME dimensions that are used as input for Enlil (see Section~\ref{subsec:default}). Specifically, we construct a circle around the CME's elliptical cross-section to account for the fact that, whilst CMEs are usually assumed to be croissant-shaped, the shocks driven by them are generally interpreted as spheroids \citep[e.g.,][]{liu2019b, dumbovic2020}. This results in a patch centred at ($\theta$, $\phi$) = ($9^{\circ}$, $5^{\circ}$)  in Stonyhurst coordinates (i.e., the propagation direction of the CME nose derived from white-light observations) and with a $49^{\circ}$ radius. To approximate either the solar flare or coronal shock source in SEPMOD, SEPs are introduced into the Enlil simulation volume at the 21.5\,$R_{\odot}$ inner boundary. These are injected over a circular area centered on the coordinates of either the PFSS-mapped flare field line(s) described above, or the cone CME in the case of the coronal shock source. Their spatial concentrations over the circular source area are described in each case by a normalised Gaussian whose half-width is based on the radius of the area. We note that the coronal source option is particularly amenable to real-time modelling because it uses the same CME source parameters employed to run the WSA--Enlil model.

Next, we need to define the SEP injection times and temporal profiles, together with their fluxes and energy spectra. For the solar flare source, we use as the SEP injection time the peak time of the associated GOES X-ray emission, i.e.\ 06:38~UT \citep[we note that][reported a radio Type~III emission starting at 06:31~UT, which could be used as an alternative constraint]{jebaraj2023}, and employ an exponential decay time that \editone{is scaled with the flare class, under the assumption that this parameter is related to the duration of the X-ray emission---e.g., \citet{kahler2022} showed that stronger flares tend to decay over longer time scales, possibly due to the longer duration of the related reconnection processes.} For the coronal shock source, we use the onset time of the corresponding radio Type~II emission \citep[considered a signature of a formed CME-driven shock; e.g.,][]{vrsnak2008, magdalenic2010}, i.e.\ 06:33~UT according to the analysis performed by \citet{jebaraj2023} employing ground-based observations. In this case we also use an exponential decay time profile, but with a time constant calculated from the transit time of the CME through the corona to the Enlil inner boundary---we remark that this choice has no physical basis, but allows for a smooth transition from the coronal shock source to the default interplanetary one. For the injected energy spectrum in the solar flare source we use a soft power law consistent with observed flare events with index $\alpha = -4.0$, while for \editone{the coronal} shock source we use a time-dependent `rollover spectrum’ with a rollover energy of 10~MeV \citep[see, e.g.,][]{desai2016} that starts with sharply diminished low energy fluxes and evolves over the time it would take the CME to arrive at 21.5\,$R_{\odot}$ to a power law with index $\alpha = -2.0$. 

\begin{figure}[t!]
\begin{center}
\includegraphics[width=.99\textwidth]{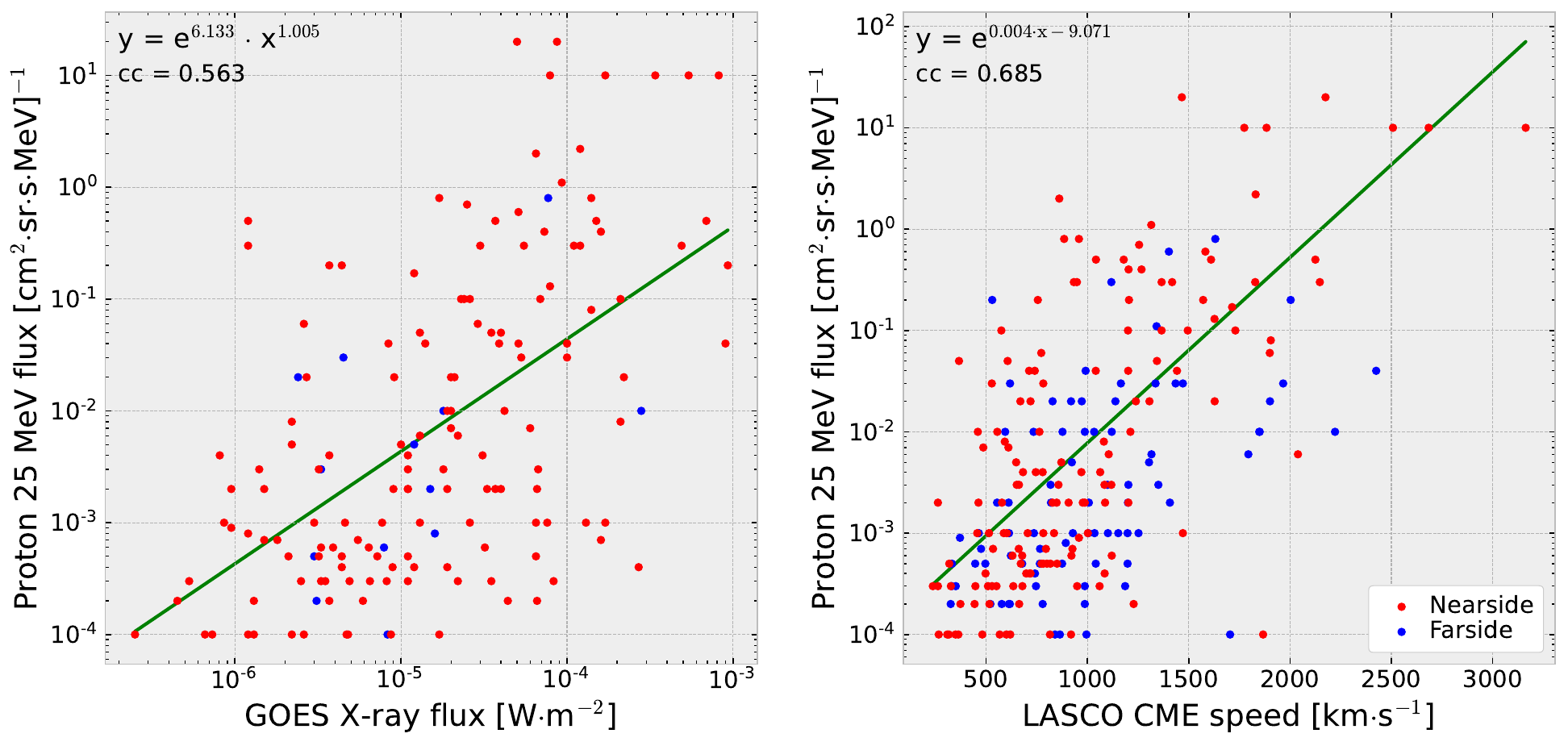}
\end{center}
\caption{Empirical relationships between the maximum proton flux at 25~MeV and (left) the GOES X-ray flux as well as (right) the LASCO plane-of-sky CME speed for SEP events detected at Earth between December 2006 and February 2018. The plot distinguishes between near-sided (red) and far-sided (blue) events. The best fit curves shown in green are applied to the near-sided data only. The top left corner of both panels shows the corresponding equation for the best fit curve and the Pearson correlation coefficient.}
\label{fig:sep_empirical}
\end{figure}

To scale the intensities of the particles injected with either the solar flare or the coronal shock sources, we use empirical relations based on SEP events detected at Earth between December 2006 and February 2018 \citep[see][for a similar approach used in operational SEP forecasting]{balch1999, balch2008}. The list of events that we use to do so is taken (and expanded) from \citet{richardson2014}, who catalogued all the $>$25~MeV proton events observed by the twin STEREO spacecraft and/or Earth. These relations are used to scale the SEP intensity to the GOES flare class for the solar flare source and to the CME plane-of-sky speed in LASCO \citep[taken, in our case, from the LASCO CME catalogue of][]{gopalswamy2009} for the coronal shock source. We first select all the events that were observed at Earth, irrespective of their detection at either STEREO probe. We then consider their peak 25~MeV flux against the corresponding GOES X-ray peak flux and LASCO CME plane-of-sky speed, and ultimately derive linear fits for both cases, shown in Figure~\ref{fig:sep_empirical}. Note that the data points shown in the figure include both near- and far-sided events, but near-sided events only were used to \editone{obtain} the linear fitting curves. The relation that we obtain for the \editone{solar} flare source option (scaled with the GOES X-ray peak flux) is
\begin{equation}
F = e^{6.133} \cdot X^{1.005} \;\;\; \mathrm{or} \;\;\; \log(F) = 1.005 \log(X) + 2.664 \, ,
\end{equation}
where $F$ is the 25~MeV proton flux and $X$ is the X-ray flux. For the coronal \editone{shock} source option (scaled with the LASCO plane-of-sky CME speed), we obtain
\begin{equation}
F = e^{0.004 V - 9.071} \;\;\; \mathrm{or} \;\;\; \log(F) = \frac{V}{\editone{576}} - 3.94 \, ,
\end{equation}
where $V$ is the CME speed. The resulting correlation coefficients (also reported in Figure~\ref{fig:sep_empirical}) are 0.563 for the flare class and 0.685 for the CME speed. We remark, \editone{nevertheless, that observational comparisons of SEP fluxes with flare class and/or CME speed like the ones we use here are not necessarily representing strictly flare-accelerated and shock-accelerated particles, respectively, but may instead present contributions from both acceleration mechanisms within a single event.}

\section{Modelling results and comparison with observations} \label{sec:result}

In this section, we compare modelling results using the various WSA--Enlil--SEPMOD setups described in Section~\ref{sec:model} with in-situ observations at the five locations depicted in Figure~\ref{fig:orbits}, i.e.\ Earth as well as the BepiColombo \citep[Bepi;][]{benkhoff2021}, Solar Orbiter \citep[SolO;][]{muller2020}, Parker Solar Probe \citep[PSP;][]{fox2016}, and STEREO-A spacecraft. These four probes were respectively located at approximately 0.33, 0.68, 0.77, and 0.96~au from the Sun at the time of the 2021 October 9 event, and their longitudinal separations from Earth were respectively 2$^{\circ}$, 15$^{\circ}$, 48$^{\circ}$, and 39$^{\circ}$ towards the east. At Bepi, we use magnetic field data from the Mercury Planetary Orbiter Magnetometer \citep[MPO-MAG;][]{heyner2021} and particle data from the BepiColombo Environment Radiation Monitor \citep[BERM;][]{pinto2022}---the plasma instruments are not operational during cruise phase. At SolO, we use magnetic field data from the Magnetometer \citep[MAG;][]{horbury2020}, plasma data from the Proton and Alpha Sensor (PAS) of the Solar Wind Analyser \citep[SWA;][]{owen2020}, and particle data from the Energetic Particle Detector \citep[EPD;][]{rodriguezpacheco2020}. At PSP, we use magnetic field data from FIELDS \citep{bale2016}, plasma data from the Solar Probe Cup \citep[SPC;][]{case2020} part of the Solar Wind Electrons Alphas and Protons \citep[SWEAP;][]{kasper2016} suite, and particle data from the Integrated Science Investigation of the Sun \citep[IS$\odot$IS;][]{mccomas2016}. At STEREO-A, we use magnetic field data from the Magnetic Field Experiment \citep[MFE;][]{acuna2008} part of the In situ Measurements of Particles And CME Transients \citep[IMPACT;][]{luhmann2008} suite, plasma data from the Plasma and Suprathermal Ion Composition investigation \citep[PLASTIC;][]{galvin2008}, and particle data from the Low Energy Telescope \citep[LET;][]{mewaldt2008} and High Energy Telescope \citep[HET;][]{vonrosenvinge2008} part of IMPACT. Finally, for Earth we employ measurements from two probes that are located at the Sun--Earth Lagrange L1 point, i.e.\ Wind \citep{ogilvie1997} and SOHO. We use magnetic field data from the Magnetic Field Investigation \citep[MFI;][]{lepping1995} as well as plasma data from the Solar Wind Experiment \citep[SWE;][]{ogilvie1995}, both onboard Wind, and particle data from the Comprehensive Suprathermal and Energetic Particle Analyser \citep[COSTEP;][]{mullermellin1995} onboard SOHO.

\subsection{The baseline: The interplanetary shock source} \label{subsec:ip_shock}

\begin{figure}[th!]
\centering
\includegraphics[width=.99\linewidth]{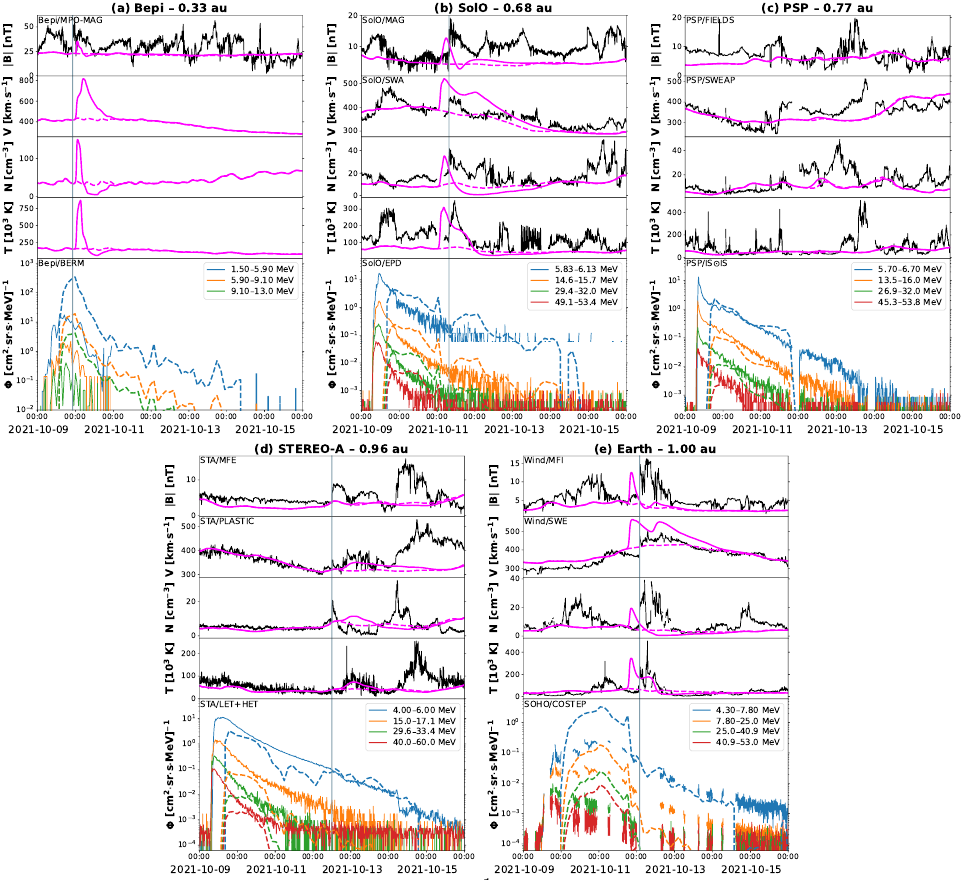}
\caption{Results of the default WSA--Enlil--SEPMOD simulation run (i.e., using the interplanetary shock source only) compared with observations of the 2021 October 9 CME and SEP event at five locations in the inner heliosphere---(a) Bepi, (b) SolO, (c) PSP, (d) STEREO-A, and (e) Earth. Each plot shows, from top to bottom: Magnetic field magntitude, solar wind bulk speed, proton density, proton temperature, and proton/ion fluxes. Magnetic field and plasma data are shown in solid black, whilst the corresponding simulated quantities are shown in solid magenta---the related ambient run (i.e., without CME) is shown in dashed magenta. Energetic particle data are shown with solid lines, whilst the corresponding simulated fluxes are shown with dashed lines. The (observed) arrival of the CME-driven shock is marked with a vertical grey line, where applicable. The heliocentric distances reported for each observer refer to the eruption time of the 2021 October 9 event.}
\label{fig:ip_shock_source}
\end{figure}

Before we can evaluate the possible improvements in modelling the onset time of SEP events with the WSA--Enlil--SEPMOD architecture, we shall explore the results obtained with the default simulation setup, i.e.\ using uniquely the interplanetary shock source (see Section~\ref{subsec:default}). An overview of magnetic field, plasma, and particle data compared to the modelling output at each of the five available observers is presented in Figure~\ref{fig:ip_shock_source} (also, see Figure~\ref{fig:enlil_setup} for a visualisation of the configuration of the various spacecraft with respect to the CME). Regarding the CME impacts at the different locations, we note that all encounters are correctly predicted, albeit with more or less significant errors in arrival time. At Bepi (Figure~\ref{fig:ip_shock_source}(a)), a shock is detected in the WSA--Enlil simulation ${\sim}$2~hours later than observed. At SolO (Figure~\ref{fig:ip_shock_source}(b)), the shock passage takes place in the model ${\sim}$5~hours earlier than in the data. At PSP (Figure~\ref{fig:ip_shock_source}(c)), there is no clear indication of a CME impact in the model, in agreement with the in-situ measurements. At STEREO-A (Figure~\ref{fig:ip_shock_source}(d)), a glancing encounter with the eastern flank of the shock is simulated ${\sim}$6~hours later than observed. Finally, at Earth (Figure~\ref{fig:ip_shock_source}(e)), the modelled CME-driven shock makes an impact ${\sim}$7~hours earlier than in the in-situ observations. Nevertheless, we remark that all impact times fall within the typical CME arrival time uncertainties, which are of the order of ${\sim}$10~hours \citep[e.g.,][]{riley2018, wold2018, paouris2021}.

Regarding the simulated SEPs at each location of interest, we note that both the intensity values and the intensity--time profiles are generally well reproduced. Despite the discrepancy in the onset times, there is a remarkable qualitative agreement between the observed and modelled SEP time series at SolO, PSP, and STEREO-A. At Bepi, the simulated fluxes are higher in magnitude than the measured ones, but the shapes of the SEP enhancements are overall well-behaved. At Earth, in addition to predicting higher fluxes during the first half of the event, SEPMOD features less steep profiles than observed, especially at higher energies---this is not surprising, since Earth is the location with the poorest magnetic connectivity to the eruption's source (see Figure~\ref{fig:orbits}). Despite the predominantly encouraging agreement between simulation and observations, it is clear that the onset of the SEP event is entirely missed at every location, with delays ranging from ${\sim}6$~hours (Bepi) to  ${\sim}11$~hours (Earth). This issue is addressed in the following sections via the fixed-source option of SEPMOD.

\subsection{Additional SEP injection: The solar flare source} \label{subsec:sol_flare}

\begin{figure}[th!]
\centering
\includegraphics[width=.99\linewidth]{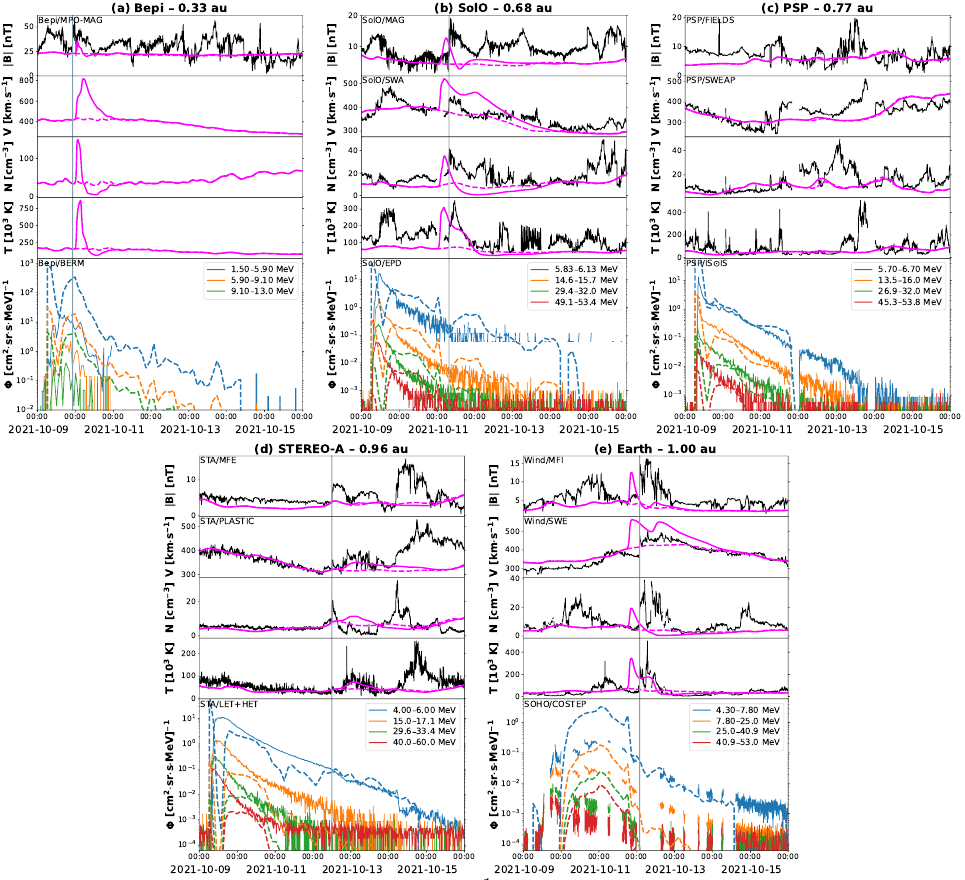}
\caption{Results of the WSA--Enlil--SEPMOD simulation run that includes the solar flare source compared with observations of the 2021 October 9 CME and SEP event at five locations in the inner heliosphere---(a) Bepi, (b) SolO, (c) PSP, (d) STEREO-A, and (e) Earth. All panels are shown in the same format as Figure~\ref{fig:ip_shock_source}.}
\label{fig:sol_flare_source}
\end{figure}

The first fixed-source option that we explore is the solar flare one. Figure~\ref{fig:sol_flare_source} (shown in the same format as Figure~\ref{fig:ip_shock_source}) displays the modelled SEP fluxes that we obtain by including, alongside the ``default'' interplanetary shock source, the additional particle injection at the Enlil inner boundary using the patch shown in Figure~\ref{fig:fixed_sources}(a). It is clear that at all locations except Earth, the onset time of the SEP event is greatly improved overall, but we note a stark ``dip'' marking the decay of the solar flare source and the takeover of the interplanetary shock source. This is a result of the assumption that the particle intensity decay is loosely based on the short lifetime of the observed 2021 October 9 flare (beginning around 06:30~UT and lasting ${\sim}$3.5~hours in total) compared to the injection time of the CME itself into the Enlil heliospheric domain beginning at 21.5\,$R_{\odot}$ (set to around 11:00~UT in the simulation run described in Section~\ref{subsec:default}).

At Bepi (Figure~\ref{fig:sol_flare_source}(a)), the simulated SEP profiles show a better onset time compared to the default SEPMOD version, but the peak fluxes---already higher-than-observed when using the interplanetary shock source only---are significantly stronger than in the corresponding measurements (even by two orders of magnitude). At SolO (Figure~\ref{fig:sol_flare_source}(b)), the modelled SEP fluxes reach their peak a couple of hours earlier than observed, and the offset between the measured and simulated flux values shortly after onset is seen to increase with increasing energy. At PSP (Figure~\ref{fig:sol_flare_source}(c)), the short-lived, quick-rising SEPMOD profiles feature a remarkably similar behaviour to the correponding measurements, although the peak fluxes at the different energy channels are more ``spread out'' in intensity than observed. At STEREO-A (Figure~\ref{fig:sol_flare_source}(d)), the modelled SEPs peak in flux earlier than observed (as in the SolO case), and the dip between the solar flare and interplanetary shock contributions is especially prominent. Finally, at Earth (Figure~\ref{fig:sol_flare_source}(e)), the fixed-source component of the simulated SEP profiles is extremely minor---only lower-energy particles exceed the instrument background---and can virtually be considered negligible.

\subsection{Additional SEP injection: The coronal shock source} \label{subsec:cor_shock}

\begin{figure}[th!]
\centering
\includegraphics[width=.99\linewidth]{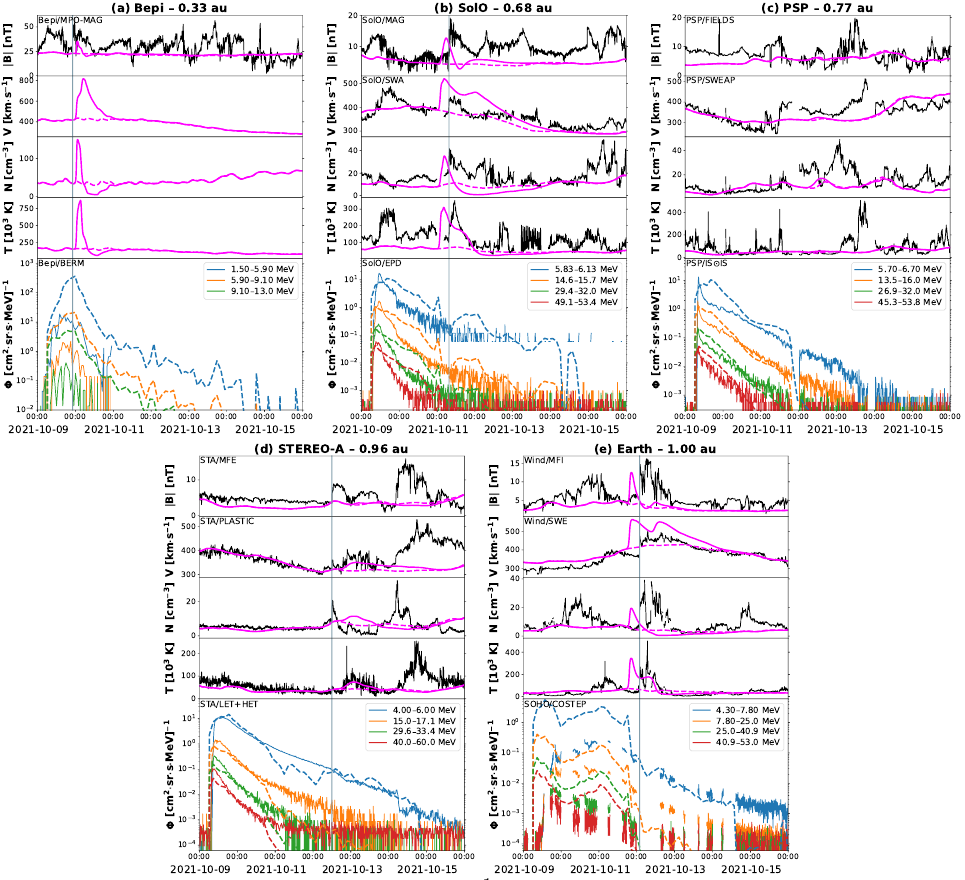}
\caption{Results of the WSA--Enlil--SEPMOD simulation run that includes the coronal shock source compared with observations of the 2021 October 9 CME and SEP event at five locations in the inner heliosphere---(a) Bepi, (b) SolO, (c) PSP, (d) STEREO-A, and (e) Earth. All panels are shown in the same format as Figure~\ref{fig:ip_shock_source}.}
\label{fig:cor_shock_source}
\end{figure}

We now investigate how results change when considering the coronal shock source. Figure~\ref{fig:cor_shock_source} (shown in the same format as Figure~\ref{fig:ip_shock_source}) displays the modelled SEP fluxes that we obtain by including, alongside the ``default'' interplanetary shock source, the additional particle injection at the Enlil inner boundary using the patch shown in Figure~\ref{fig:fixed_sources}(b). Again, the onset time of the SEP event appears greatly improved overall, and this time we note a significantly smoother transition between the coronal and interplanetary shock sources, resulting in a more realistic time series. This is likely due to the fact that the coronal shock source is based on the same CME input parameters that are employed for Enlil, and its decay time profile is scaled with the CME transit time to 21.5\,$R_{\odot}$.

At Bepi (Figure~\ref{fig:cor_shock_source}(a)), the modelled SEP profiles display a similar trend to the measured ones, but all the magnitudes are higher in all cases (as in the previous two simulation runs). At SolO (Figure~\ref{fig:cor_shock_source}(b)), PSP (Figure~\ref{fig:cor_shock_source}(c)), and STEREO-A (Figure~\ref{fig:cor_shock_source}(d)), the modelled fluxes appear remarkably similar to the observed ones, in terms of both the shapes of the SEP enhancements and in the peak intensities at the different energy levels. The SEPMOD fluxes drop below the corresponding instrumental backgrounds earlier than in the spacecraft data in a few cases, including the two lowest energy ranges at PSP (5.70--6.70 MeV and 13.5--16.0 MeV) as well as the two mid energy levels at STEREO-A (15.0--17.1 MeV and 29.6--33.4 MeV). Finally, at Earth (Figure~\ref{fig:cor_shock_source}(e)), SEPMOD simulates a significantly more abrupt onset of the SEP event than observed, and with generally higher fluxes at all energy ranges.


\section{Discussion} \label{sec:discuss}

The results presented in Section~\ref{sec:result} show that the SEPMOD fixed-source option that we have implemented and tested on the 2021 October 9 SEP event yields overall earlier onset times---generally closer to the observed ones---as well as greater peak fluxes over different energy ranges---often more similar to the corresponding measurements, but significantly higher in a few instances. Here, we discuss such results (especially in terms of the heliospheric context) and attempt to quantify the level of improvement of both the solar flare and coronal shock sources over the default interplanetary shock source.

First of all, we note that the solar flare source, despite producing overall more accurate onset times than the default model and partially better peak fluxes, yields SEP profiles that are characterised by an abrupt, short-lived peak followed by a dip and, only after, by the default interplanetary shock contribution. As mentioned in Section~\ref{subsec:sol_flare}, this is due to the short SEP injection time that is scaled to the corresponding X-ray flare emission. An exception is found, in part, at PSP, where an initial short-lived peak---without, however, a prominent dip following---is present in the spacecraft data as well. This is, in fact, the location that is best connected to the eruption's source region (see Figure~\ref{fig:orbits}). It is possible that PSP observed the SEP event with both a flare and a CME-driven shock contribution \citep[e.g.,][]{cane2010, anastasiadis2019}. The coronal shock source, on the other hand, produces better onset times and more realistic profiles overall. As mentioned in Section~\ref{subsec:cor_shock}, this is likely due to the fact that this source is scaled to the CME parameters used for the Enlil run---and from which SEPMOD derives its default interplanetary shock source---thus yielding a more seamless transition between the two SEP injection modes. An exception is found at Earth, where the coronal shock contribution results in a more rapid onset and higher peak fluxes. This is the location characterised by the poorest connectivity to the eruption's source region (see Figure~\ref{fig:orbits}). It is likely that the patch selected for the coronal shock SEP injection was too large at the beginning of the event, since CME-driven shocks are usually formed in the lower corona and tend to expand rapidly during their early propagation \citep[e.g.,][]{kwon2014, liu2019a}. Future improvements to the SEPMOD coronal shock source will consider a linearly-growing patch from the first CME observation time in coronagraphs until its ``maximum'' size used as input for Enlil.

\begin{table}[t!]
\caption{Differences in the SEPMOD modelled onset times of the 2021 October 9 SEP event at the various observers using the default code and the two fixed-source options. \label{tab:onsettimes} \vspace*{.1in}}
\centering
\renewcommand{\arraystretch}{1.3}
\begin{tabular}{l@{\hskip .6in}c@{\hskip .6in}c@{\hskip .4in}c@{\hskip .3in}c}
\toprule
 & \textsc{\textbf{Observed}} & \textsc{\textbf{Default}} & \textsc{\textbf{+ Solar Flare}} & \textsc{\textbf{+ Coronal Shock}} \\
\midrule
\textsc{\textbf{Bepi}} & 08:00~UT & +6.0~h & $-$1.5~h & $-$1.5~h \\
\textsc{\textbf{SolO}} & 07:30~UT & +6.5~h & $-$1.0~h & $-$1.0~h \\
\textsc{\textbf{PSP}} & 07:00~UT & +7.5~h & $-$0.5~h & $-$0.5~h\\
\textsc{\textbf{STEREO-A}} & 07:30~UT & +6.5~h & $-$1.0~h & $-$1.0~h \\
\textsc{\textbf{Earth}} & 13:00~UT & +11.0~h & +11.0~h & $-$6.5~h \\
\bottomrule
\end{tabular}
\vspace*{.1in}
\begin{tablenotes}
\item \emph{Notes.} The first column shows the approximate (to the nearest half-hour) observed SEP event onset times at each spacecraft, whilst the remaining columns report the corresponding differences in the simulated onset times for different versions of SEPMOD. Positive (negative) offsets indicate that the predicted onset took place later (earlier) than observed. For each time estimate, we have referred mainly to the orange SEP time series in Figures~\ref{fig:ip_shock_source}, \ref{fig:sol_flare_source}, and \ref{fig:cor_shock_source}, with energy ranges of 5.9--9.1~MeV at Bepi, 14.6--15.7~MeV at SolO, 13.5--16.0~MeV at PSP, 15.0--17.1~MeV at STEREO-A, and 7.8--25~MeV at SOHO.
\end{tablenotes}
\end{table}

Table~\ref{tab:onsettimes} provides an overview of the different SEP onset times modelled by SEPMOD compared to their corresponding observations. For each observer, we report the actual event onset times (approximated to the nearest half-hour) as well as the deviation from those for each of the SEPMOD runs employed in this work. It is clear that the default (interplanetary shock source) version yields times that are delayed at least by a few hours in all cases, as mentioned in Section~\ref{subsec:default}. Both fixed-source options, on the other hand, bring overall significant improvements, with onset times that are 30 to 60~minutes earlier than observed at all locations but Earth. The fact that these SEP events start earlier than in the corresponding spacecraft data may be due to either a larger path length along the interplanetary  magnetic field than that modelled by Enlil, or an SEP injection that was set to commence too soon in SEPMOD (note that we used here the flare peak time for the solar flare source and the onset of the Type~II radio emission for the coronal shock source). At Earth, the solar flare source produces no significant changes when compared to the default simulation run---likely due to the source patch being too small for particles to reach this location---whilst the coronal shock source yields an onset time that is significantly earlier than observed---likely due to the source patch being too large during the early phases of SEP acceleration, as mentioned above.

Table~\ref{tab:maxfluxes} provides an overview of the different SEP peak fluxes modelled by SEPMOD compared to their corresponding observations. For each observer, we report the measured SEP intensity peak in one of the energy channels shown in Figures~\ref{fig:ip_shock_source}, \ref{fig:sol_flare_source}, and \ref{fig:cor_shock_source} as well as the corresponding value for each of the SEPMOD runs employed in this work. It is clear that at SolO, PSP, and STEREO-A, the default version of SEPMOD tends to significantly underestimate the peak fluxes. This is not surprising, since for well-connected observers an SEP event usually reaches its highest intensities close to onset. The solar flare source, however, shows still rather low fluxes at SolO and STEREO-A, whilst at PSP intensities are overestimated by approximately a factor of two. The coronal shock source, on the other hand, produces SEP peak fluxes at these three observers that are no worse than ${\sim}$0.6 times the measured ones. At Bepi and Earth, all three SEPMOD simulation runs produce fluxes that are higher than observed. Whilst at Bepi this could be due to the packed configuration of the instrument during its cruise phase, implying that the spacecraft itself may screen some particles \citep[see also Appendix~A in][]{palmerio2022}, it is clear that at Earth this is owing to either a too-large SEP injection patch or the specific heliospheric context---\citet{lario2022} and \citet{wijsen2023}, in fact, showed that there was a solar wind high-speed stream flowing between the CME source region and Earth's location, which may have resulted in particles being partially occulted and/or delayed.

\begin{table}[ht!]
\caption{Differences in the SEPMOD modelled peak intensities of the 2021 October 9 SEP event at the various observers using the default code and the two fixed-source options. \label{tab:maxfluxes} \vspace*{.1in}}
\centering
\renewcommand{\arraystretch}{1.3}
\begin{tabular}{l@{\hskip .6in}c@{\hskip .5in}c@{\hskip .4in}c@{\hskip .3in}c}
\toprule
 & \textsc{\textbf{Observed}} & \textsc{\textbf{Default}} & \textsc{\textbf{+ Solar Flare}} & \textsc{\textbf{+ Coronal Shock}} \\
\midrule
\textsc{\textbf{Bepi}} & 2.232 & 19.123 (857\%) & 32.308 (1447\%) & 21.487 (963\%) \\
\textsc{\textbf{SolO}} & 1.734 & 0.245 (14\%) & 0.371 (21\%) & 1.069 (62\%) \\
\textsc{\textbf{PSP}} & 1.963 & 0.116 (6\%) & 4.159 (212\%) & 1.202 (61\%) \\
\textsc{\textbf{STEREO-A}} & 1.471 & 0.081 (6\%) & 0.289 (20\%) & 0.862 (59\%) \\
\textsc{\textbf{Earth}} & 0.047 & 0.184 (391\%) & 0.184 (391\%) & 0.411 (874\%) \\
\bottomrule
\end{tabular}
\vspace*{.1in}
\begin{tablenotes}
\item \emph{Notes.} The reported fluxes refer to the orange SEP time series in Figures~\ref{fig:ip_shock_source}, \ref{fig:sol_flare_source}, and \ref{fig:cor_shock_source}, with energy ranges of 5.9--9.1~MeV at Bepi, 14.6--15.7~MeV at SolO, 13.5--16.0~MeV at PSP, 15.0--17.1~MeV at STEREO-A, and 7.8--25~MeV at SOHO. The units are [cm$^{2}\cdot$sr$\cdot$s$\cdot$MeV]$^{-1}$ for all reported values. The percentages in parentheses for the SEPMOD runs are calculated with respect to the corresponding observed fluxes.
\end{tablenotes}
\end{table}

Finally, to provide a comparison of the different modelled SEP fluxes versus their corresponding spacecraft data as a whole, we employ the Dynamic Time Warping \citep[DTW;][]{bellman1959} technique, which is an algorithm useful for measuring similarity between two time series. In DTW, an optimal comparison (more specifically, alignment) between two data sequences is found by stretching and/or compressing one series onto the other, and the resulting DTW distance is a measure of their similarity---the lower the distance, the higher the correlation. The technique has found applications in many fields of science and technology, including more recently in heliophysics \citep[e.g.,][]{laperre2020, samara2022}. To apply the DTW method to our data sets, we first ensure that any pair of time series covers the same temporal interval, and successively linearly interpolate any data gaps that may be present in the spacecraft measurements. Table~\ref{tab:dtwdist} shows the DTW distances resulting from comparing observations of the 2021 October 9 SEP event with each of the three SEPMOD runs considered in this work. Note that the values reported in the table are not normalised, i.e., they should not be compared across the different in-situ observers for each SEPMOD running mode, but only across the different SEPMOD runs for each spacecraft. It is clear that for SolO, PSP, and STEREO-A, the coronal shock source is characterised by a significantly lower (by at least a factor of two or three) DTW distance than the remaining two, indicating a better match to observations. For all these three observers, the solar flare source also displays consistently lower values than the default (interplanetary shock only) option. In the case of Bepi, the default run provides the smallest DTW distance, followed by the coronal shock and then by the solar flare sources---we note, nevertheless, that the best and worst values differ by less than a factor of two. At Earth, the interplanetary shock and the solar flare sources display equal results, whilst the coronal shock source features a DTW distance that is ${\sim}$2.5~times higher. When considering all these results as a whole, it appears that the coronal shock source (and, to a lesser extent, the solar flare one) represents overall an improvement over the default SEPMOD running mode, which neglects SEP acceleration below heliocentric heights of 21.5\,$R_{\odot}$.

\begin{table}[t!]
\caption{DTW distances between spacecraft data and SEPMOD profiles of the 2021 October 9 SEP event at the various observers using the default code and the two fixed-source options. \label{tab:dtwdist} \vspace*{.1in}}
\centering
\renewcommand{\arraystretch}{1.3}
\begin{tabular}{l@{\hskip .6in}c@{\hskip .6in}c@{\hskip .4in}c}
\toprule
 & \textsc{\textbf{Default}} & \textsc{\textbf{+ Solar Flare}} & \textsc{\textbf{+ Coronal Shock}} \\
\midrule
\textsc{\textbf{Bepi}} & 26.931 & 43.597 & 35.714 \\
\textsc{\textbf{SolO}} & 4.593 & 4.003 & 1.451 \\
\textsc{\textbf{PSP}} & 3.357 & 3.303 & 0.891 \\
\textsc{\textbf{STEREO-A}} & 6.301 & 4.917 & 1.853 \\
\textsc{\textbf{Earth}} & 0.282 & 0.282 & 0.682 \\
\bottomrule
\end{tabular}
\vspace*{.1in}
\begin{tablenotes}
\item \emph{Notes.} The calculated DTW distances refer to the orange SEP time series in Figures~\ref{fig:ip_shock_source}, \ref{fig:sol_flare_source}, and \ref{fig:cor_shock_source}, with energy ranges of 5.9--9.1~MeV at Bepi, 14.6--15.7~MeV at SolO, 13.5--16.0~MeV at PSP, 15.0--17.1~MeV at STEREO-A, and 7.8--25~MeV at SOHO. The units are [cm$^{2}\cdot$sr$\cdot$s$\cdot$MeV]$^{-1}$ for all reported values.
\end{tablenotes}
\end{table}


\section{Summary and conclusions} \label{sec:conclu}

In this work, we have implemented the so-called fixed-source option in the SEP transport code SEPMOD, and applied it within the framework of MHD heliospheric modelling using the WSA--Enlil--SEPMOD architecture. The reasoning behind this effort is to improve modelling of the onset and early evolution of SEP events, which often cannot be reproduced in models that include CME transients only in their heliospheric domain (often set at 21.5\,$R_{\odot}$ or 0.1~au). We have employed two separate fixed SEP sources, namely a solar flare (with SEP properties modulated by the corresponding flare characteristics) and a coronal shock (with SEP properties modulated by the corresponding CME characteristics), aimed at emulating the two sites where particle acceleration may occur. We have tested both these sources on the 2021 October 9 SEP event, which was triggered by an eruption characterised by an M1.6-class flare and a moderate-speed (${\sim}$770~km$\cdot$s$^{-1}$) CME, and found a series of improvements over the default SEPMOD running mode, which only includes particle acceleration at the CME-driven interplanetary shock above 21.5\,$R_{\odot}$.

Overall, results obtained with the coronal shock source appeared to match observations better than the other two simulation modes in terms of event onset timing, peak fluxes, and similarity with the corresponding spacecraft measurements. This is not surprising, given that the 2021 October 9 event was characterised indeed by a strong shock-accelerated component \citep[see][for details on the corresponding ESP event, showing that the shock was still accelerating particles by the time it reached 1~au]{wijsen2023}, hence the solar flare source option may not have been particularly appropriate for this case. In the context of real-time space weather forecasting, the coronal shock source may be in general a simpler option to implement, since it uses directly the CME parameters employed for the related MHD simulation of the solar wind and interplanetary magnetic field (in this case, with WSA--Enlil), whilst the solar flare source requires the additional step of performing a PFSS extrapolation of the nearby open fields. Additionally, the properties of the flare source that are relevant for characterizing the SEP event are generally less clear. For example, we have assumed that the particle intensity rate is loosely dependent on the X-ray flare decay time, which may not be correct, used a soft SEP spectrum based on small flare events that may not be well-suited for these more energetic events, and assumed that the flare-accelerated particles are released over a small region of open fields based on a PFSS extrapolation, which may not be appropriate where the field configuration is strongly evolving during the eruption of a CME. Flare acceleration by reconnection may also occur over a wider region beneath the CME assumed for the PFSS extrapolation. Another caveat that holds true for either fixed-source option is that, as mentioned in Section~\ref{sec:model}, SEPMOD neglects cross-field SEP transport (at least in its current implementation), hence these assumptions and results may be generally applicable only to events for which each location of interest is relatively well-connected to the eruption's source region and/or the CME-driven shock, as was the case for the 2021 October 9 event explored here.

With these caveats in mind, in future work we will consider \editone{more} stricly-impulsive SEP events accompanied \editone{e.g.\ by jets or narrow/slow CMEs} \citep[e.g.,][]{nitta2006, papaioannou2016} and test the solar flare source option further. Additional future efforts will consider employing a linearly growing source patch for the coronal shock source, to emulate the expanding behaviour of CME-driven shocks in the corona (and to avoid SEPs reaching poorly-connected observers too early, as was the case for Earth in this work). In fact, the implementation of a time-dependent coronal shock source could be combined with the solar flare source description introduced here to obtain realistic-looking SEP profiles that are characterised by both a flare- and a shock-accelerated component (as was likely the case for PSP in this work)---such an option shall also be explored in the future. Additionally, we note that these developments to the SEPMOD code were tested and applied thoroughly uniquely on the event showcased in this work. A more systematic survey of many more events observed by multiple spacecraft in the inner heliosphere is underway and will reveal the applicability of the fixed-source description presented here for any given SEP-producing eruption.

\editone{Nevertheless, we remark that SEPMOD is a simplified-physics model that assumes scatter-free propagation of SEPs along magnetic field lines, whilst neglecting additional particle transport processes \citep[e.g.,][]{giacalone2001, qin2004}. In fact, effects such as diffusive/perpendicular transport \citep[e.g.,][]{zhang2003, zank2015}, which have been shown to often play a significant role in the spread of SEPs \citep[e.g.,][]{laitinen2013, dresing2014}, are not included in the current implementation of SEPMOD. We have described in this section a set of potential improvements to our semi-empirical treatment of the flare and/or coronal shock sources, but additional avenues to pursue could include the incorporation of cross-field transport to yield a more realistic description of the spread of particles in the heliosphere. Since one of the major strengths of SEPMOD is its computational efficiency, which makes the model particularly well-suited for real-time applications\footnote{See, e.g., the SEP Scoreboard hosted at NASA's Community Coordinated Modeling Center (CCMC), available at \href{https://ccmc.gsfc.nasa.gov/scoreboards/sep}{https://ccmc.gsfc.nasa.gov/scoreboards/sep}, which regularly includes SEPMOD predictions in real-time applications.}, one of the greatest challenges in this regard is the attainment of an optimal balance between physics-based formulations and semi-empirical approximations that can provide sufficient physical insight whilst maintaining simulation runs executable over short time scales.}

\editone{Finally, it is worth keeping in mind that simulating the acceleration and transport of SEPs with architectures such as the one employed in this work will necessarily involve inherent uncertainties that go beyond the mere treatment of particles and are related to various aspects of the full Sun-to-heliosphere modelling chain. These include uncertainties in the photospheric magnetic field measurements \citep[e.g.,][]{bertello2014, riley2014}, in the reconstructed coronal fields \citep[e.g.,][]{caplan2021, parenti2021}, in the modelled ambient solar wind \citep[e.g.,][]{gressl2014, jian2015}, and in the estimation of CME input parameters \citep[e.g.,][]{singh2022, verbeke2023}. In other words, the parameter space of the simulation has a large number of poorly-constrained degrees of freedom, thus making generalisation of the method for space weather forecasting applications particularly challenging. Extensive efforts dedicated to quantitative comparisons and performance benchmarking for each of these pieces of the modelling chain \citep[e.g.,][]{reiss2023b, reiss2023a, temmer2023, whitman2023} will be increasingly important for model refinement and development and advancing our space weather prediction capabilities.}


\begin{acknowledgements}
EP and RMC acknowledge support from NASA's PSP-GI (no.\ 80NSSC22K0349), O2R (no.\ 80NSSC20K0285), LWS (no.\ 80NSSC19K0067), and LWS-SC (no.\ 80NSSC22K0893) programmes.
DL and IGR acknowlege support from NASA programmes NNH19ZDA001N-LWS and
NNH19ZDA001N-HSR. IGR also acknowledges support from the STEREO mission.
COL acknowledges support from NASA LWS grant 80NSSC21K1325.
BS-C acknowledges support through STFC Ernest Rutherford Fellowship ST/V004115/1 and STFC grant ST/Y000439/1.
NW acknowledges funding from the Research Foundation -- Flanders (FWO -- Vlaanderen, fellowship no.\ 1184319N).
DH is supported by the German Ministerium f{\"u}r Wirtschaft und Klimaschutz and the German Zentrum f{\"u}r Luft- und Raumfahrt under contract 50QW2202.
The authors thank NASA's Community Coordinated Modeling Center (CCMC; \href{https://ccmc.gsfc.nasa.gov}{https://ccmc.gsfc.nasa.gov}) for supporting the simulation efforts presented in this work (run id: \textsl{Erika\_Palmerio\_111522\_SH\_1}). The WSA model was developed by C.~N.~Arge (currently at NASA Goddard Space Flight Center), the Enlil model was developed by D.~Odstrcil (currently at George Mason University), and the SEPMOD model was developed by J.~G.~Luhmann (currently at University of California--Berkeley).
\end{acknowledgements}

\bibliography{bibliography}
   

\end{document}